\documentclass[11pt]{article}

\usepackage{epsf}

\newcommand{\beq}{\begin{equation}}
\newcommand{\eeq}{\end{equation}}
\newcommand{\beqa}{\begin{eqnarray}}
\newcommand{\eeqa}{\end{eqnarray}}

\def\a{\alpha}
\def\b{\beta}

\def\t{\tau}

\def\del{\partial}
\def\rb{\bar{r}}
\def\rs{{R^\ast}}
\def\rmm{r_{\mbox{\scriptsize max}}}
\def\rbah{\bar{r}_{\mbox{\scriptsize ah}}}
\def\tc{\t_{\mbox{\scriptsize crash}}}
\def\fot{\frac{1}{2}}
\def\ftt{\frac{3}{2}}

\begin{document}

\begin{flushright}
gr-qc/9608050
\\
August 20, 1996
\end{flushright}

\vfill

\begin{center}{\Large\bf
Adaptive mesh and \\[4mm]
geodesically sliced Schwarzschild spacetime \\[4mm] 
in 3+1 dimensions
}
\end{center}
\vfill

\begin{center}
\large Bernd Br\"ugmann
\end{center}

\begin{center}
{\em Max-Planck-Institut f\"ur Gravitationsphysik,
 \\ Schlaatzweg 1, 14473 Potsdam, Germany \\[2mm]}
{\tt bruegman@aei-potsdam.mpg.de}
\end{center}

\vfill

\begin{center}
\large\bf Abstract
\end{center}
\medskip

\noindent
We present first results obtained with a 3+1 dimensional adaptive mesh
code in numerical general relativity. The adaptive mesh is used in
conjunction with a standard ADM code for the evolution of a
dynamically sliced Schwarzschild spacetime (geodesic slicing). We
argue that adaptive mesh is particularly natural in the context of
general relativity, where apart from adaptive mesh refinement for
numerical efficiency one may want to use the built in flexibility to
do numerical relativity on coordinate patches.

\vspace*{\fill}

\noindent
PACS numbers: 04.25.Dm, 95.30.Sf, 97.60.Lf

\newpage

\section{Introduction}

One of the stepping stones towards unrestricted 3+1 dimensional
numerical general relativity is the study of the Schwarzschild space
time. Spacetime singularities are one of the two characteristic
features of vacuum general relativity, the other being gravitational
waves. We certainly have to learn how to deal with a single, static
black hole numerically if we want to treat astrophysically more
interesting scenarios like the collision of two black holes, the final
stage of which is again a single static black hole.

The static, spherically symmetric Schwarzschild space time turns into a
rather challenging test case for the standard 3+1 dimensional numerical 
evolution schemes, if one does not make use of the spherical symmetry
other than in the initial data, if one uses cartesian coordinates, and
if one uses the freedom in the 3+1 decomposition to define
hypersurfaces on which the metric components evolve in time. 
This is what we implement here, following closely the work of Anninos
et\ al. on 3+1 \cite{a}, 
which in turn is based on Bernstein et\ al. on 1+1 \cite{b} (by $n+1$
we denote the use of one time and $n$ space variables). 

`Adaptive mesh' refers to a general technic for numerical evolution
problems based on discrete grids, the basic idea being that one puts
the points where one needs them for a given numerical accuracy.  While
traditionally the domain of numerical computation is taken to be a
single, fixed rectangular grid (with several field variables per
point), the suggestion is to monitor the numerical errors, and
wherever and whenever the error becomes too big, an additional finer
grid is introduced. Similarly, if the error is small enough, the grids
are adjusted and possibly removed altogether. Since the error is
changing dynamically, this results in a dynamically changing structure
of several levels of nested grids.

Of course, the idea to adapt the resolution to the observed numerical
error has a long history, and is now commonplace in many areas of
numerical computation. For solving initial value problems for ODEs, there are
Runge-Kutta methods with adaptive step-size control, or the
Bulirsch-Stoer algorithm \cite{NuRe}.  Multigrid methods were promoted
already in the seventies by Brandt for the solution of elliptic
problems in any number of dimensions \cite{Bra}. 
For hyperbolic systems the basic
reference is the work by Berger and Oliger on adaptive mesh refinement
\cite{BeOl}. In the context of general relativity, adaptive mesh
refinement has been made famous by Choptuiks pioneering work on the
collapse of a spherically symmetric scalar field \cite{Ch:am,Ch:prl}.
In order to resolve all the details of the Choptuik effect in 1+1, a
refinement by a factor on the order of $10^7$ over the initial
resolution is required. Clearly, if one was to repeat these
calculations for more than one spatial variable, the efficiency of
adaptive mesh becomes essential.

Perhaps it is appropriate to ask at this point why adaptive mesh ---
which is such an obvious and simple idea --- is not in widespread use in
general relativity. There are two main reasons. 

First, one does have to be able to solve the equations of interest on
a uniform grid. This turns out to be a rather hard problem in general
relativity, where only a handful of codes in 3+1 has been developed
\cite{handful} due to general relativistic problems related to
space time singularities and the choice of lapse and shift. (As of
1996, we feel it is no longer justified to list limited computer
resources as a main reason.)

Second, programming an adaptive mesh is rather complicated, and it is of a
different nature than other programming tasks in numerical relativity
because it involves dynamically changing data structures. 

In this paper we address both these problems. Based on prior
experience with dynamical data structures (in dynamical triangulations
for Monte Carlo simulations in four dimensional Euclidean quantum
gravity \cite{BrMa}), it was not too difficult to implement an
adaptive mesh code in two and three spatial dimensions. The code was
tested as an empty adaptive mesh for a given error function, and for
the scalar wave equation in flat space.

As a concrete test case in general relativity, we settled on a
3+1 Schwarzschild spacetime in geodesic slicing.
Let us emphasize that this is not a showcase for the capabilities of
our adaptive mesh code, only up to three nested grids are
involved. But having a general adaptive mesh package available allowed
us to automatically use a coarse grid in the outer regions and finer
grids near the interior of the black hole. The gain in efficiency in
turn allowed us to perform computations on a small workstation that
compare well with those that the NCSA group performed on various
supercomputers \cite{a}.   

Let us spell out briefly what constitutes the core of our test runs.
Given appropriate initial data for the Schwarzschild spacetime at the
moment of time symmetry in spatially isotropic, cartesian coordinates
$x$, $y$, and $z$, the evolution in time $t$ is computed using the
standard ADM equations and an explicit finite difference scheme
(double leapfrog), where we choose lapse $\a \equiv 1$ and shift $\b^a
\equiv 0$, which induces geodesic slicing.  A point starting at
initial Schwarzschild radius $r = 2M$ reaches the singularity at $ r=
0$ after proper time $\t = \pi M$, where $M$ is the mass of the black
hole.  We also evolve the data up to about $\t = 6M$ by using the
apparent horizon as the inner boundary \cite{SeSu}. The resulting dynamical
evolution of, for example, the six metric coefficients can be directly
compared to the analytic solution.

The adaptive mesh code can, to a certain extent, be thought of as a
black box. The user has to supply just one external input, a routine
that evolves data on a uniform grid with a given boundary.  While the
outer boundary can be incorporated easily into this routine, for the
case of an apparent horizon boundary condition it was simpler to
customize the adaptive mesh itself, i.e. to incorporate grids with
`holes'.

The author is aware of two other adaptive mesh refinement packages
that are currently under development and that are planned to be
applied to 3+1 numerical relativity: DAGH of the American grand
challenge collaboration \cite{DAGH}, and a code by Wild \cite{ScWi}.
It is interesting to note that the problem independent design of DAGH
does not include grids with holes, but for reasons similar to ours
that are specific to general relativity, this feature will be added.

Finally, we want to draw attention to how naturally adaptive mesh fits
into general relativity. While the numerical point of view leads us to
drive the adaptivity of the adaptive mesh technic by the numerical
errors, general relativity gives us a physical reason to split the
domain of computation into several grids, namely simply that one of
the main characteristics of general relativity is that space time is a
manifold, which generically can only be covered by several charts, and
which can be covered by charts in which the metric is almost flat.
This leads us to discuss `numerical relativity on
patches' below. In fact, some of the features that make geodesic
slicing inattractive for numerical relativity may loose their impact
when combined with adaptive mesh. To underscore our point of view we
deviate from common terminology and use the term `adaptive mesh' as
opposed to the less general `adaptive mesh refinement'.

The paper is organized as follows. In section 2, we introduce various
coordinate systems for the Schwarzschild space time and the standard
3+1 decomposition. In section 3, we describe our uniform ADM code. In
section 4, we discuss some issues related to adaptive mesh in general,
while in section 5, we present our particular implementation. In
section 6, we discuss results obtained for adaptive mesh and
Schwarzschild space time in geodesic slicing. We conclude with section
7.

\section{Schwarzschild space time and geodesic slicing}

The line element for a single static black hole in Schwarzschild
coordinates is given by
\beq
	ds^2 = - (1 - \frac{2M}{r}) dt^2 + (1-\frac{2M}{r})^{-1} dr^2
+ r^2 d\Omega^2,
\eeq
where $M$ is the mass, $r$ the radius, and $d\Omega^2$ the standard
line element on the unit 2-sphere. We define spatially isotropic
coordinates by introducing a new radial coordinate $\rb$, such that
\beqa
	r &=& \rb (1+\frac{M}{2\rb})^2, 
\label{rrb}
\\
	ds^2 &=& - \a(\rb)^2 dt^2 + \psi(\rb)^4 (d\rb^2 + \rb^2
        d\Omega^2), 
\\
	\a(\rb) &=& (1-\frac{M}{2\rb})/(1+\frac{M}{2\rb}), 
\label{isolapse}
\\
	\psi(\rb) &=& 1+\frac{M}{2\rb}.
\label{psi}
\eeqa
This allows us to introduce the cartesian spatial coordinates that we
use in the numerical computations,
\beq
	dx^2 + dy^2 + dz^2 = d\rb^2 + \rb^2 d\Omega^2, 
	\quad \rb = (x^2+y^2+z^2)^{1/2}.
\eeq
The spatially isotropic coordinates possess an isometry at the throat
at $\rb = M/2$ for $\rb \leftrightarrow \rb' = M^2/(4\rb)$, e.g.\
$r(\rb) = r(\rb')$ and $\a(\rb) = - \a(\rb')$. The isotropic
coordinates for $\rb\in
[M/2,\infty]$ and $\rb\in[M/2,0]$ cover the same range of the
Schwarzschild radius, $r \in [2M,\infty]$.

In the standard 3+1 decomposition of the Einstein equations (e.g.\
\cite{Yo}), the line element can be written in general as
\beq
	ds^2 = - (\a^2 - \b^a\b_a) dt^2 + 2 \b_a dt dx^a + g_{ab} dx^a
dx^b,
\eeq
where $\a$ is the lapse function, $\b^a$ the shift vector, and
$g_{ab}$ the three-metric. The Einstein equations decompose into
the Hamiltonian and diffeomorphism constraint equations, and the
evolution equations for the $g_{ab}$ and their canonically conjugate
momenta, the extrinsic curvature $K_{ab}$,
\beqa
	\del_t g_{ab} &=& -2\a K_{ab} + D_a\b_b + D_b\b_a, 
\label{dgdt}
\\
	\del_t K_{ab} &=& -D_aD_b\a + \a (R_{ab} + K_{ab} {K^c}_c - 2
K_{ac}{K^c}_b) \nonumber 
\\
	&& + \b^cD_cK_{ab} + K_{ac} D_b\b^c + K_{cb}D_a\b^c,
\label{dKdt}
\eeqa
where $R_{ab}$ is the 3-Ricci tensor, and $D_a$ the covariant
derivative defined for the 3-metric. 

The generic evolution problem is: given some initial data for $g_{ab}$
and $K_{ab}$ (solving constraints), and a prescription for $\a$ and
$\b$, and boundary conditions, construct the space time. We make the
following choices. For coordinates $t$, $x$, $y$, and $z$, we define
the initial 3-metric at $t=0$ by
\beq
	\mbox{}^{(3)}ds^2 = \psi(\rb)^4 (dx^2+dy^2+dz^2),
\label{ginit}
\eeq
where the conformal factor $\psi$ is defined in (\ref{psi}).  The
initial data for the extrinsic curvature is determined by making $t=0$
the moment of time symmetry, $K_{ab} = 0$. This initial data is a
solution to the constraints.

There are several methods to fix the freedom in the definition of the
3+1 decomposition, and making a good choice is essential because
otherwise the evolution will break down due to physical or coordinate
singularities after a short time. In particular, a lot of work has
been carried out on singularity avoiding slicing conditions (e.g.\
\cite{a} and references therein). Here we choose geodesic slicing,
$\a\equiv1$ and $\b^a\equiv0$, so that points with constant spatial
coordinates follow geodesics and $t$ becomes the proper time. The
initial data corresponds to observers or test particles that are
initially at rest and then start falling towards the singularity (no
singularity avoidance). For a discussion of problems related to
geodesic slicing, see section 4. To test our code we also checked that
choosing vanishing shift and the lapse of the quasi isotropic
coordinates, (\ref{isolapse}), the configuration does not change
(which considering (\ref{dKdt}) is a nontrivial numerical problem).

We now have to specify the boundary conditions. As the outer
boundary we consider the limit in which $\rb \rightarrow \infty$. In
general, there does not exist something like a `purely outgoing wave
condition' at finite radius for non-linear equations like the Einstein
equations, because in general purely outgoing waves are not an exact
solution (there always is back scattering). Some approximation is
usually the simplest way to proceed, and in our case, similar to
\cite{a}, it is sufficient to set all fields equal to their initial
value at the outer boundary, as long as it is located at $\rb$
sufficiently large. More elaborate procedures are certainly possible,
but in conjunction with adaptive mesh not necessary for our problem,
since adaptive mesh allows us to go out to sufficiently large values
of $\rb$.

We define an inner boundary for intermediate $\rb$ by either using the
isometry at the throat, $\rb = M/2$ \cite{a}, or by cutting off the
spacetime at the horizon, $r = 2M$ \cite{SeSu,ADMSS}. In the former
case, the isometry defines a simple coordinate transformation from
which one can compute the values of the fields for $\rb < M/2$ once
the fields are known for $\rb > M/2$. Note that $\rb = M/2$ refers to
an unchanging location in our coordinates, but $r = 2M$ defines a
curve $\rb = \rbah(\t)$ for the location of the (apparent)
horizon. The apparent horizon boundary condition derives from the fact
that the horizon is a null surface, so that the exterior is causally
disconnected from the interior.

As in \cite{a}, to reduce the computational effort by a factor of 8,
most computations are carried out on the octant of positive $x$, $y$,
$z$ only, and the reflection symmetry of spherical symmetry at the $x
= 0$, $y=0$, and $z=0$ planes is used to derive boundary values via a
simple coordinate transformation. We did check the code also on the
full grid, and it seems quite unlikely that enforcing symmetry on
these planes only suffices to ensure spherical symmetry everywhere.

Given the precise evolution problem just stated, what do we know about
the resulting space time? A convenient feature of geodesic slicing is
that the result can be directly compared to the analytic solution. It
is somewhat amusing to note that the two previous numerical papers on
the topic do not make use of the well-known analytic solution, but in
\cite{b} on 1+1 the validity of the numerical results is established
mostly from internal consistency (apart from the crash test), and
\cite{a} on 3+1 check their results against \cite{b}. 
Of course, in general it is much more useful to be able to check a
code without having the analytical solution available, but since it
happens to be available in this case, we use it here.

Unit lapse and vanishing shift define Gaussian normal coordinates,
which in the context of the Schwarzschild space time are called
Novikov coordinates \cite{No,MTW}. These are the comoving coordinates
in which radially moving freely falling test particles are at rest and
the time coordinate measures proper time.  Starting from Schwarzschild
coordinates, there are several natural coordinate transformations. One
can find a transformation to spatially isotropic coordinates, or
to unit light cones (Kruskal), or to proper time, but of course not
simultaneously to proper time and spatial isotropy.

In Schwarzschild coordinates, a
radial geodesic starts at $r = 0$ and performs a cycloidal motion
out to some maximal radius $\rmm$ and back to $r = 0$.
The Schwarzschild geometry in Novikov coordinates is given in terms of
a new radial coordinate $\rs$ by
\beqa
	\rs &=& (\frac{\rmm}{2M} - 1)^\fot, 
\label{rsrm}
\\
	ds^2 &=& -d\t^2 + \frac{\rs^2 +1}{\rs^2} (\frac{\del
r}{\del\rs})^2 d\rs^2 + r^2 d\Omega^2,
\eeqa
where $r = r(\t,\rs)$ is implicitly given by the following relation
obtained from integrating the geodesic equation,
\beqa
	\frac{\t}{2M} &=& \pm (\rs^2 +1) (\frac{r}{2M} -
\frac{(r/(2M))^2} {\rs^2 + 1})^\fot \nonumber \\
        && + (\rs^2 + 1)^\frac{3}{2} \arccos (
(\frac{r/(2M)}{\rs^2+1})^\fot).
\label{trrs}
\eeqa
To actually compute $r(\t,\rs)$ we have to invert a relation of the
type $y = x + \sin(x)$, which can only be done numerically, but in a
very simple manner (e.g.\ by bisection). 

An important property of Gaussian normal coordinates is that the
geodesics that define the coordinates remain orthogonal to all
constant time hypersurfaces. Therefore, the coordinate transformation
between $\rb$ and $\rs$ obtained by inserting (\ref{rrb}) into
(\ref{rsrm}) is time independent. On the other hand, since $r$ is a
function of time, the data does not remain isotropic.

To explicitly compute interesting quantities like the metric
coefficients for $M = 1$, $\rs > 0$, and $\t >0$, we find it
convenient to use the maximal Schwarzschild radial coordinate $\rmm$,
for which
\beqa
	\rmm &=& 2(\rs^2 + 1) = \frac{(1+2\rb)^2}{4\rb},
\\
	\t &=& \rmm (\frac{r}{2}(1 - \frac{r}{\rmm}))^\fot + 2
(\frac{\rmm}{2})^\ftt \arccos ((\frac{r}{\rmm})^\fot),
\label{trrm}
\eeqa
and by implicit differentiation,
\beq
        \frac{\del r}{\del \rmm} = \ftt - \frac{r}{2\rmm} + 
	\ftt (\frac{\rmm}{r} -1) \arccos ((\frac{r}{\rmm})^\fot).
\eeq
For example, transforming from $\rs$ to $\rb$ leads to a simple
formula for the radial metric component,
\beq
        g_{\rb\rb}(\t,\rb) = \psi(\rb)^4 (\frac{\del r}{\del \rmm})^2
	(r(\t,\rmm(\rb)), \rmm(\rb)),
\label{grbrb}
\eeq
where as before $g_{\rb\rb}$ depends on time through $r$, which is
given implicitly by (\ref{trrm}) as $r(\t,\rmm)$.  Considering that
the (time independent) conformal factor $\psi(\rb) = 1 + M/(2\rb)$
diverges at $\rb = 0$, it is natural to compute $g_{ab}/\psi^4$ to
focus on the dynamical features in the metric rather than on the
static $1/\rb$ singularity, as is done in \cite{a} and as we
often do below. Equation (\ref{grbrb}) justifies this approach.

\begin{figure}
\epsfxsize=15cm
\centerline{\epsffile{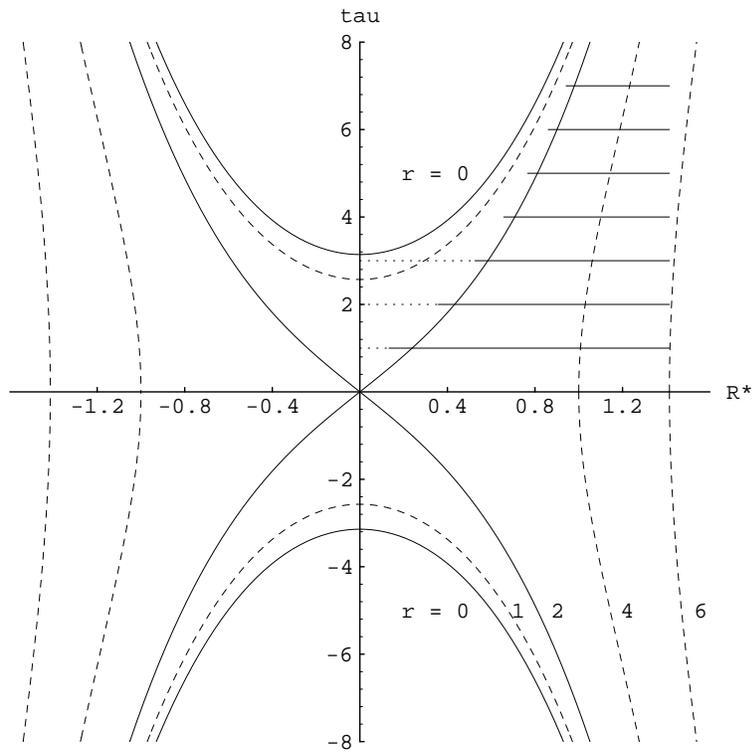}}
\caption{Novikov coordinates for the Schwarzschild geometry.}
\label{nov}
\end{figure}

Figure \ref{nov} shows a plot of lines of constant $r$ based on
(\ref{trrs}) to depict the Schwarzschild geometry in Novikov
coordinates (compare with the qualitative picture in
\cite{MTW}). Note that from (\ref{trrs}) we have for the horizon
$\t/2M\approx\rs^3$ for large $\t$, as opposed to Kruskal coordinates
in which the horizon is a unit light cone. The horizontal lines show
the location of grids with and without apparent horizon boundary
condition.

An initial grid at $\t = 0$ covering $2M\leq r \leq r_0$, 
or, equivalently, $0\leq\rs\leq\rs_0$ or $M/2\leq\rb\leq\rb_0$,
moves upwards until the innermost point that started on the horizon
reaches the $r = 0$ singularity after time
\beq
	\tc = \pi M.
\eeq
In numerical `crash tests' \cite{a,b} one indeed finds for this scenario
that $\tc = (3.1 \pm 0.2) M$ \cite{a}, and our 3+1
code reproduces this result. One can also track how the radial metric
component $g_{\rb\rb}(\t,\rb)$ (constructed from $g_{ab}$) diverges
with time with an exploding peak developing at the throat at $\rb =
M/2$ \cite{a,b}. Indeed, from the analytic solution we find for $\t = 3
M$ that at the throat $g_{\rb\rb} = 20.486$ compared to $20.2$ in 1+1
and $23.4$ in 3+1 \cite{a}. 

As a test of our numerical code and in order to present some novel
data, we plot $g_{\rb\rb}(\t,\rbah(\t))$, i.e.\ how the radial metric
component develops with time, in section 6. Having summarized the
analytic aspects and some of the numerical history of the Schwarzschild
space time in Gaussian normal coordinates, we now discuss the actual
implementation of our code.

\section{ADM code for uniform grids}

The evolution equations (\ref{dgdt}) and (\ref{dKdt}) for $g_{ab}$ and
$K_{ab}$ are rather elegant and concise, but since the indices run
from 1 to 3, writing out each term explicitly leads to a problem of a
size that makes the use of computer algebra highly recommended, if not
essential for the added flexibility. We wrote a simple Mathematica
script that takes equations (\ref{dgdt}) and (\ref{dKdt}) directly as
an input, together with formulas for lapse and shift, and also some
control quantities like the constraints, translates the derivatives
into finite differences, and outputs C code for the basic routine that
evolves data on a uniform grid. A typical implementation leads to
about 1520 summations, 969 multiplications, and 322 divisions for 18
basic fields.

We choose to perform an unconstrained evolution using explicit finite
difference schemes. The schemes tested are Lax-Wendroff, double
leapfrog, and Brailovskaya, with and without artificial dissipation (see
\cite{b} for a comparison of schemes). As far as data storage is
concerned, only the double leapfrog scheme really requires two and not
one level of preceeding data, that is, the same field at two earlier
times. Although this is an additional complication for the adaptive
mesh code, we implemented it in order not to introduce a
limitation. Most production runs are performed with the double
leapfrog scheme. (Compare with \cite{a} where a particular version of
staggered leapfrog with extrapolation for the inhomogeneous terms is
used.)  Second order spatial derivatives are differenced symmetrically
with centered differences, which seems to maintain spherical symmetry
rather well, although from experience with the Laplace operator and
elliptic equations one might expect that some asymmetric differencing
is a better choice (e.g.\ \cite{MiGr}).

In \cite{a}, it was observed that for a stable evolution it was
crucial to perform differencing of the scaled metric $g_{ab}/\psi^4$,
which Anninos et al.\ called conformal differencing. We also had to
apply this technic. Since one might argue that the generality of the
evolution scheme is compromised by building in knowledge about the
initial data (recall that at least $\psi$ does not change with time),
let us add a few comments. Clearly, approximating the limiting $1/\rb$
dependence with finite differences, i.e. essentially with polynomials,
is problematic. But as a matter of principle, there always is the
issue whether an approximation method works in a given function
space. To reduce the problem dependence of conformal differencing, we
tested a somewhat more generic method, where a given type of test
function, e.g. a rational function, is fit to the data. The result of
the fit is used as a basis for `scaled' differencing. For a perfect
fit, one is left with finite differencing a constant. For the problem
at hand, however, the simple conformal rescaling was quite sufficient.

As already discussed, at the inner boundary we put either the isometry
condition or the apparent horizon boundary condition \cite{ADMSS}. Not
only the field values near the boundary that are needed for the finite
difference molecules, but all points in the interior can be obtained
by the isometry map from the data that was evolved outside. Since
interior points are in general mapped to points falling between the
outer grid points, a polynomial interpolation is performed taking due
care near the border that no data is accessed before it is available.

For the apparent horizon condition we have postponed the
implementation of a general 3+1 apparent horizon finder, and simply
define a surface by the equation $r = 2M$ leading to $\rb =
\rbah(\t)$. Following \cite{ADMSS} on 1+1, we evolve everywhere outside 
$\rbah(\t)$ minus some small buffer zone. At various times the inner
most points are obtained by second or third order polynomial (or
rational function) extrapolation. The basic algorithm can deal with a
convex surface which is sufficiently flat on the scale of the grid
points. With some fine tuning, the inner boundary remains stable with
a buffer zone of about 2.8 grid spacings. In \cite{ADMSS} a minimal size
of five and recommended size of twenty grid spacings is reported.

Note that for geodesic slicing all light cones are upright, so the CFL
condition (which requires the physical domain of dependence to be
contained in the numerical domain of dependence) reduces to the
condition that the angle between the physical characteristics and the
$\t$ coordinate lines is not too big. At large $\rb$, the light cones
approach the unit cone (45 degrees for $c=1$), and it is simple to see
that near the horizon the light cones become narrower in the radial
direction but wider in the constant radius directions ($g_{\t\t} =
-1$, $g_{\rb\rb}$ increases, and for the azimuthal angle $\theta$,
$g_{\theta\theta} = r^2$ decreases).

The numerical domain of dependence is related to the physical one by
the factor by which the temporal grid spacing is smaller than the
spatial grid spacing. For the finite difference schemes considered, a
relative factor of $0.25$ was used, although a factor of $0.1$ made
the evolution slightly more accurate (but slower). We did not
encounter the problem that the light cones become too narrow or too
wide.

\section{Numerical relativity on patches}

Before getting into the details of the adaptive mesh code, we would
like to discuss a few issues related to numerical relativity and
adaptive mesh in general. As explained in the introduction, the basic
idea is to put points where they are needed for a given accuracy, but
in the general relativistic setting a more general view point is
possible.

A typical text book introduction to general relativity may proceed as
follows. First one learns that gravitational physics is really about a
manifold with a metric. In the neighborhood of any point the manifold
looks like $R^4$, but in the generic case one needs an atlas of
coordinate patches to cover the manifold. Furthermore, there always
exist coordinates near a point, in which the metric is close to the
flat Minkowski metric. To borrow a picture from Einstein's discussion
of the principle of general covariance \cite{MTW}, consider the
gravitational field of the earth. Everywhere around the earth we can
construct freely falling frames of reference which approximate
Minkowski spacetime, but no single set of coordinates exists in which
space time is everywhere flat. So the space time structure with its
locally flat patches is a key feature of general relativity, and let
us emphasize that, apart from global (topological) issues, it is also
a key feature in a practical sense if we look for coordinates in which
the metric is locally flat.

Ironically, the next step is to completely ignore, or at least,
circumvent the patch work character of general relativity. One learns
a lot about beautiful work where a single or a few coordinate patches
are constructed in an ingenious manner to cover all the interesting
regions of spacetime \cite{Ne}. Typically this involves using special
symmetries of the model. What is perhaps more relevant for generic 3+1
dimensional numerical relativity without symmetries, for simple
initial data the original coordinate system stays good for at least
some time close to the initial hypersurface, so again one might try to
make do with one coordinate system rather than changing coordinates.

Numerical relativity has been traditionally built upon one or a few
handcrafted grids, mostly fixed for the whole evolution. There are
very well-known examples for problems associated with rigid boxes, to
name just one, the steep gradients in the metric developing for maximal
slicing of the Schwarzschild space time \cite{a}, with a promising
solution being the apparent horizon boundary condition where the grid
adapts itself to the apparent horizon and is not strictly fixed.

What we want to suggest is that the adaptive mesh technic encompasses
the necessary flexibility to actually implement numerical relativity
on patches. Namely, it may be possible to drive the automatic
distribution of grids not only by numerical error estimates, but also
by some physical measure. For example, such that the new grids
correspond to coordinate patches in which the metric is nearly flat,
or has some other convenient property like minimal distortion.

Let us emphasize that to us this suggestion appears to be of the type
nobody would object to --- as long as one can produce a concrete and
useful implementation. This is not done here, except perhaps for one
aspect discussed below. But we want to develop the idea a little bit
further in an illustrative thought experiment for the ADM formalism,
and for geodesic slicing.

\begin{figure}
\epsfxsize=10cm
\centerline{\epsffile{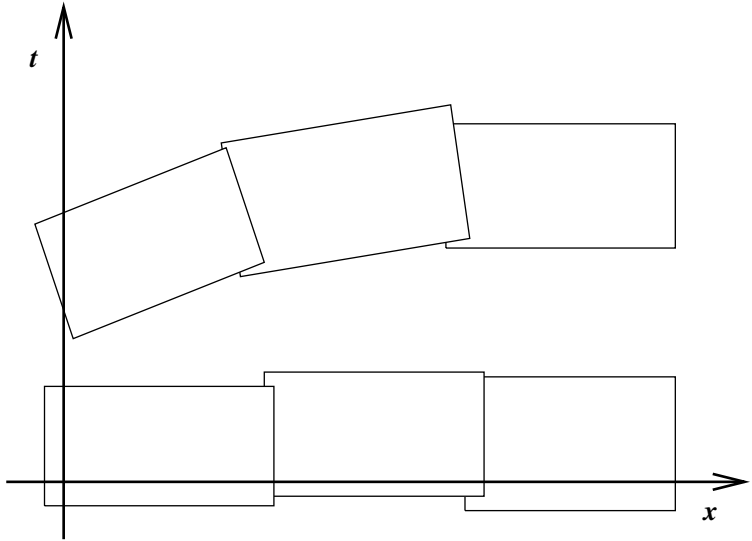}}
\caption{Schematic example for numerical relativity on patches.}
\label{amp}
\end{figure}

The main idea is displayed in figure \ref{amp}.  Suppose we are
given initial data that is well represented according to some
criterion like local flatness of the metric. For a brief time, this
criterion does not lead us to regrid, and the data evolves in the
rectangular space time patches that are drawn near the $x$-axis in
figure \ref{amp}. Now suppose that figure \ref{amp} corresponds in a
rough sense to a black hole in that in the course of evolution the
light cones are tilted inwards toward the $t$-axis, and that freely
falling observers follow an inward curving path. The flatness
criterion could lead at later times to the second row of patches,
where the initial boxes have been adapted to the inward tilt.

Several comments are in order. Note that the original work by Berger
and Oliger \cite{BeOl} already contains the concept of rotating boxes
(to track shock fronts), although only in spatial directions. Here the
suggestion is to construct boosted data by similar interpolation
technics. 

Also note that the data structures are often based on a strict nesting
property, which has been relaxed in a twofold sense. Here boxes of
equal refinement are allowed to overlap, a feature which one also
needs if areas of refinement are to be covered by several small boxes.

A more severe way in which the nesting property has to be violated is
that some internal boxes might not have a parent. This must be
allowed, since the parent box might correspond to a region of space
time which is too warped to be adequately covered by a single grid. A
concrete way to implement parentless grids is to define a transition
function which must exist since we are dealing with a manifold.

Finally note that no adaptive mesh refinement might be involved at
all, i.e.\ all grids could have the same grid spacing, although one
needs a coordinate independent measure of the grid spacing to make
this statement meaningful.

In terms of the standard 3+1 decomposition of Einstein's equations,
the above example amounts to a particular choice of both lapse and
shift. Introducing a shift vector so that all light cones are upright
is the subject of what is called causal differencing
\cite{SeSu} or causal reconnection \cite{AlSc}. 
The difference is that for the particular patches just introduced, the
causally correct differencing is discrete on the scale of the grid
sizes and not of the grid spacing. In section 3, we discussed that
even when the light cones are upright, one still has to adjust lapse
and/or temporal and spatial grid spacings. In three spatial
dimensions, upright light cones intersect the hypersurfaces in a
non-spherical manner, and on the patches one might want to define
coordinates such that the cross sections approximate spheres.

To complement this qualitative discussion of numerical
relativity on patches, let us conclude the section with
a few comments on how the transition from a single, fixed domain of
computation to varying patches might be of help for the two main
problems that are associated with geodesic slicing.
Gaussian normal coordinates have the intrinsic problem that freely
falling observers tend to fall into physical singularities, and that
coordinate singularities develop due to geodesic focussing.
 
Suppose we had some stable method to stop computing at points where
the data becomes infinite.  If all one is given is a fixed finite
grid, the grid may have to be unfeasibly large if one wants to cover a
given period of time before all points have hit the
singularity. But, considering the Schwarzschild space time in figure
\ref{nov}, even if the outermost points are far enough outside to only
move an negligible distance in the time of interest, the innermost
points fall in, leading to grid stretching near the horizon. Adaptive
mesh is helpful in this regard since it has the built-in capability to
introduce new points near the horizon.

\begin{figure}
\epsfxsize=200pt
\centerline{\epsffile{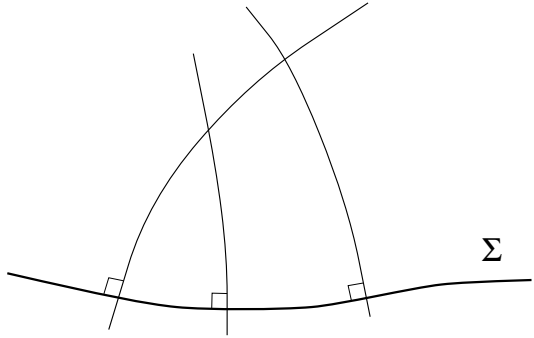}}
\caption{Geodesic focussing.}
\label{foc}
\end{figure}

For a schematic picture of geodesic focussing, consider figure
\ref{foc}. Schwarzschild space time is special since all radial 
geodesics meet at $r = 0$. The crucial point is that one has to
determine by some means, e.g.\ by evaluating curvature scalars,
whether one approaches a physical or a coordinate singularity. If
there is a physical singularity, then the adaptive mesh can insert
finer and finer grids and avoid the singularity, if we decide that
this is the feature we want to resolve rather than, for example, to
impose an apparent horizon boundary condition (if an horizon covers
the singularity). Adaptive mesh, of course, cannot change the
underlying physics.  If there is a coordinate singularity, and this is
the pathological feature of Gaussian normal coordinates we want to
address, then quantities like curvature scalars will appear more and
more constant as we approach the intersection of geodesics. The
adaptive mesh solution is to regrid, that is, to redistribute points
on a coarser grid because the finer resolution is not needed since
there is no physics to resolve.

We do not seriously want to suggest that geodesic slicing is a
universally good choice. For example, having a non-vanishing shift
vector might be crucial.  But in principle, adaptive mesh offers the
possibility to resolve the problems of geodesic slicing with its built
in capability to add in points where needed when others fell into a
physical singularity, and to remove points that otherwise would lead to
a coordinate singularity. Put the other way around, while on a fixed
grid geodesic slicing is certainly problematic, on adaptive meshes
these problems are not unavoidable. 

The numerical work of this paper can also be considered as a step
towards a demonstration that adaptive mesh can fill in points for
grids that move and stretch towards a singularity. As the horizon
moves outwards, the innermost grid expands to cover the outer regions
where points are missing to achieve the given accuracy.

\begin{figure}
\epsfxsize=15cm
\centerline{\epsffile{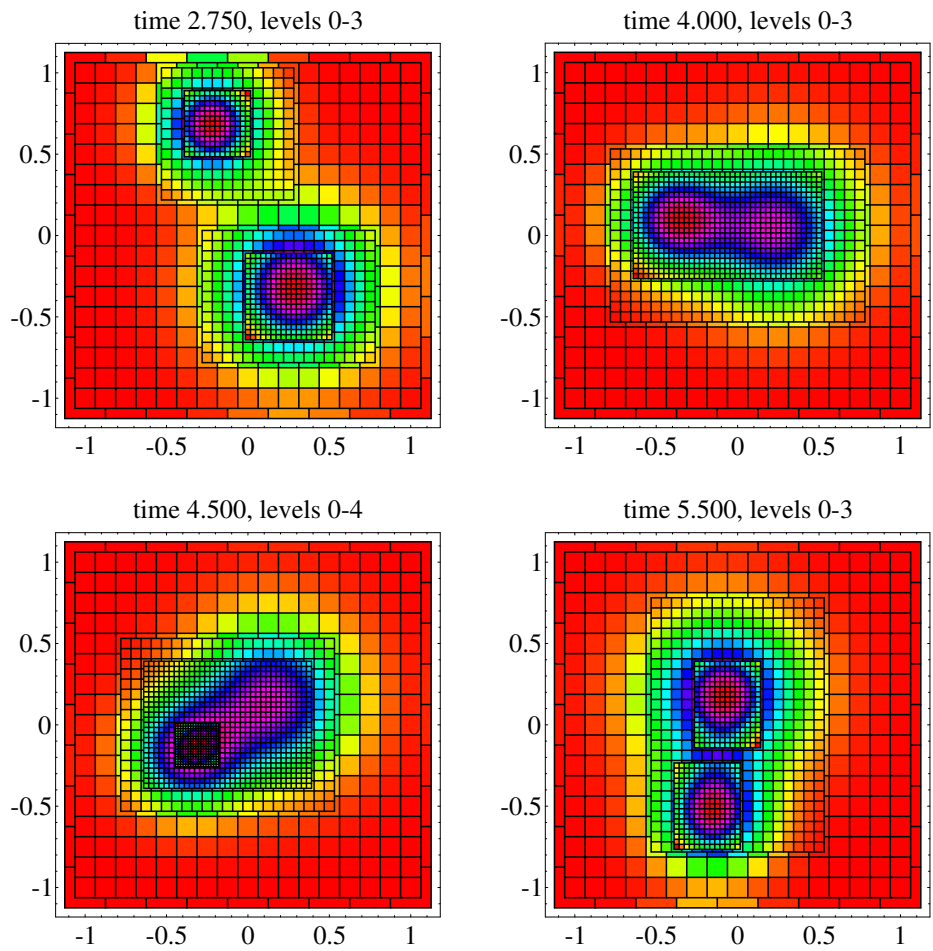}}
\caption{Empty adaptive mesh displaying various possible regriddings.}
\label{emp}
\end{figure}

\section{Implementing adaptive mesh refinement in 2 and 3 spatial
dimensions}

All current implementations of adaptive mesh in general relativity
derive from Berger and Oliger \cite{BeOl}, and are motivated and
influenced by Choptuik \cite{Ch:am}. We refer to these papers for more
technical information, but comment on important features of our
code. We should mention that there is at least one non-standard
approach, by Schutz and Wild
\cite{ScWi}, where 
the units of refinement are not grids but single points. Our main
focus is on 3+1, but some features are tested in 2+1 for simplicity.
For a visual impression of how the intuitive idea of adaptive mesh
translates into various evolving grids that follow some data, see
figure \ref{emp}, which is discussed below.

The central design issue is how to organize the dynamical data
structures. While the basic idea of structuring the grids based on
some given error estimate is very simple, it turns out to be a rather
complicated matter to have access to all the necessary information at
all times. We use linked lists of C structures describing rectangular
grids which are referenced by C pointers. This description still
leaves a lot of freedom whether one maintains pointers to all or none
of: the parent grids, the child grids, neighboring grids (we work in
three spatial dimensions), equal level grids and so forth. Depending
on the actual physics problem, it becomes a tradeoff between the cost
to maintain all these pointers versus gain in overall speed. For
simplicity, we settled on one choice without worrying about
optimization, and for 3+1 numerical relativity, most time is spent
during uniform evolution.

A very helpful idea for testing turned
out to be the concept of an empty adaptive mesh. Instead of
considering an evolution based on a differential equation, one could
consider some fake evolution, for which one also specifies a fake
error function. But all that the adaptive mesh is adapting to is the
error function, so we are considering empty adaptive meshes which
track the evolution of some predetermined error function without
reference to any data. 

Figure \ref{emp} shows two regions of error circling each other in 2+1
dimensions. This models the situation of a neutron star binary, for
which we might also expect the error to be large where the density is
high (although this is not necessarily the case). The color coding is
normalized separately for each grid to set off the subgrids. The fine
grids follow the peaks, boxes of equal refinement merge and split,
and finer boxes are inserted and removed.

One part of the adaptive mesh code is to find appropriate boxes around
volumes where the error is beyond a certain threshold, or
equivalently, to find the bounding box for flagged points among
unflagged points. To find rectangular bounding boxes, we start with a
seed and let each of its faces move outwards in turn as long as there
are flagged points on it, and since the volume grows, we have to
repeatedly consider each surface. The optimal performance is obtained
for sets of flagged points which form solid boxes, since starting with
any seed, it is a linear process to walk out to the surface, and in
order to decide that the final surface does not contain flagged
points, $O(N^2)$ operations are required for a box of volume
$N^3$. Actually, even if we had to look at each single point inside
the final box a few times, the time spend on finding boxes would be
negligible compared to on the order of $1000 N^3$ floating point
operations carried out per point during evolution. Note that this
algorithm will group disconnected regions when appropriate (e.g.\
non-convex regions whose bounding boxes overlap), which is a
big advantage over certain flood-fill algorithms.

Note also that putting an upper limit on the volume to which a seed
may grow offers a simple way to break up big regions into several
small boxes. We have not implemented this yet, but this certainly is a
good way to improve efficiency once storage for on the order of 100
reasonably sized boxes is available (as opposed to a current limit of
about 5 in 3+1). For example, the black hole space time we consider
poses the problem to cover a spherical shell, the region near the
horizon, which contains far fewer points than the bounding
box. Referring to section 4, spatially non-uniform refinement can be
useful, but since the refinement factor is constant in each grid, one
needs a larger number of boxes to adequately break up large
non-uniform regions.

One aspect of adaptive mesh that is not testable in empty adaptive
mesh by its very definition is how the error estimates are obtained.
As usual, we compute the Richardson truncation error, which involves
comparing data from the evolution on coarse and fine grids. 

Another very important issue not adressed with empty adaptive mesh is
the question of how to obtain the boundary data for the interior
subgrids.  We refine the grid spacing for both space and time by the
same factor (any integer larger than 1), so there are time steps for
which a grid is not covered by a coarser grid at equal time from which
the boundary could be interpolated. But evolving the coarser grids
first, any grid is always sandwiched between two coarser grids in
time. The coarsest grid is only allowed to have outer boundaries,
which have to be treated by different means anyway. In our examples it
has worked well to derive the boundary for the finer grids by
polynomial interpolation of order no higher than three from the two
coarser grids. We tested the interpolation first for a scalar field in
2+1 and 3+1 dimensions (planar and spherical waves) before proceeding
to the black hole case.

It is not clear whether the interior boundaries introduced by the
adaptive mesh can be treated completely independently of the evolution
scheme as we do it here with the above interpolation scheme. As
already mentioned in the introduction, we found it useful to open up
the black box concept for adaptive mesh somewhat by incorporating the
apparent horizon boundary condition as boxes with holes, even though
the apparent horizon is an outer boundary of the domain of
computation.  Also recall that we have experimented only with explicit
difference schemes for an unconstrained evolution. Whenever a
non-local operation has to be performed, for example, in an implicit
difference scheme or when solving an elliptic boundary value problem,
it is still possible to evolve the coarse grid first for the region
where the coarse data is valid. But the non-locality might introduce a
new source of noise into the system.

Regridding noise is the one additional numerical problem introduced by
adaptive mesh. Every time the grids change, there will be an
unavoidable numerical error due to interpolation and injection of
data. In our examples, a sufficiently fine grid spacing kept the
regridding noise at small enough levels.  Artificial dissipation reduces
the noise, but was not essential.

\section{Adaptive mesh and geodesically sliced Schwarzschild space
time}

\begin{figure}
\epsfxsize=15cm
\centerline{\epsffile{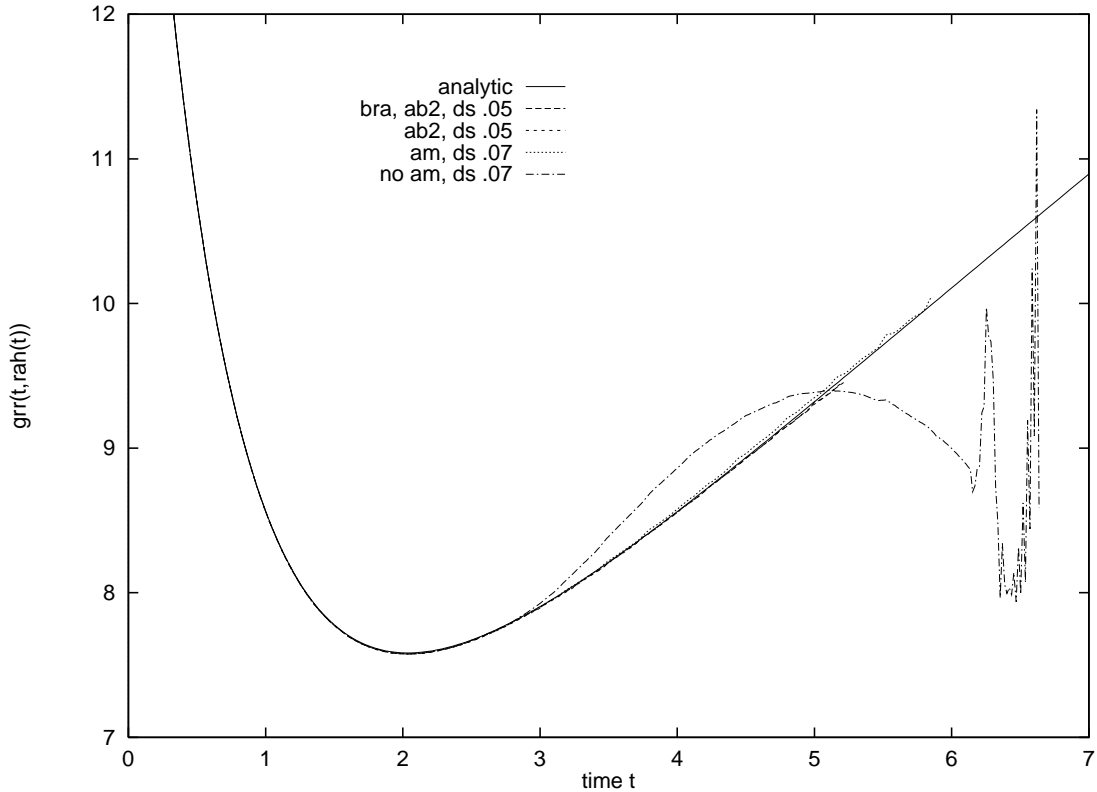}}
\caption{The metric component $g_{\rb\rb}$ at the horizon.}
\label{grr}
\end{figure}

\begin{figure}
\epsfxsize=15cm
\centerline{\epsffile{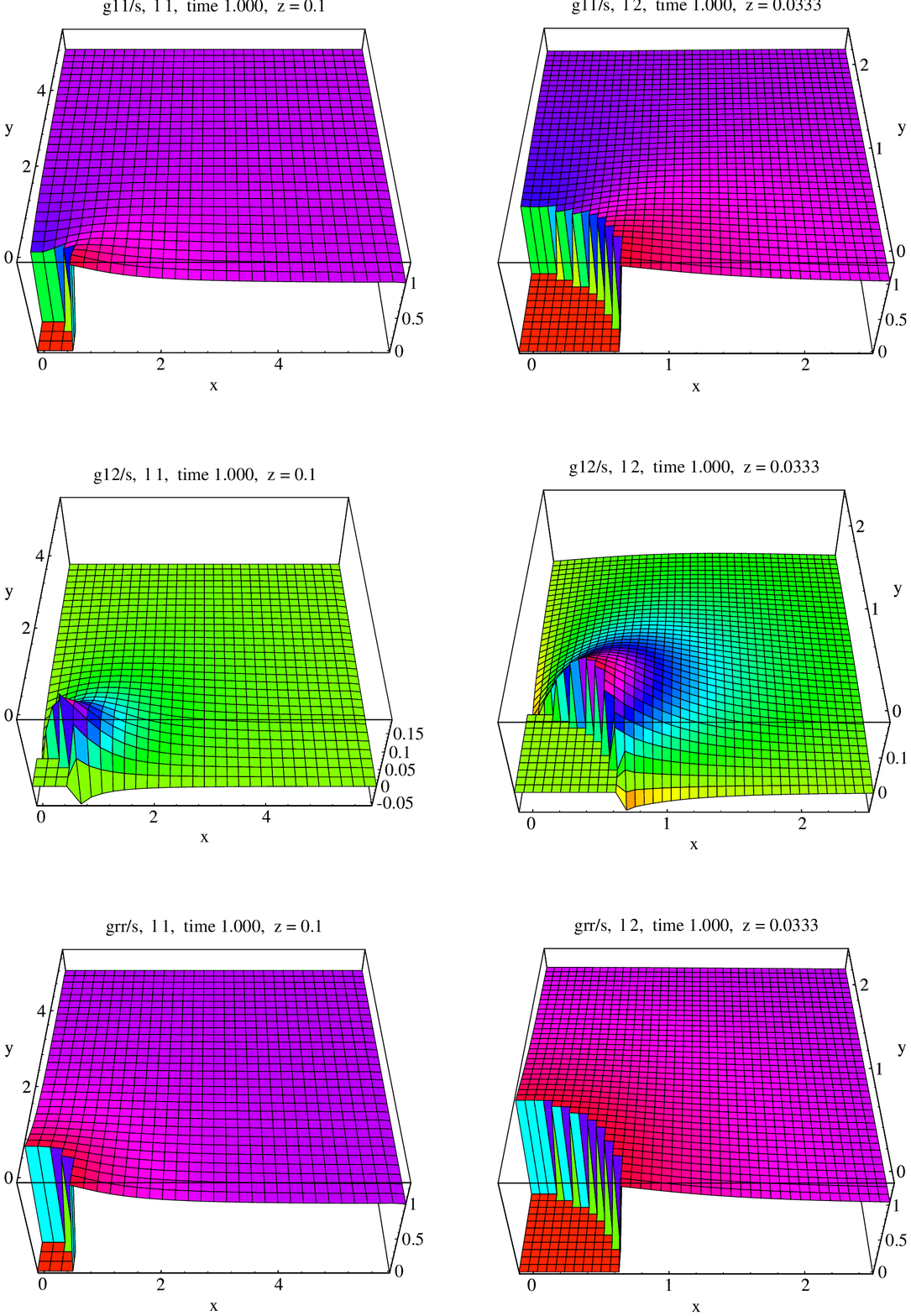}}
\caption{Metric components at levels 1 and 2 at time $\t = 1.0$.}
\label{t1}
\end{figure}

\begin{figure}
\epsfxsize=15cm
\centerline{\epsffile{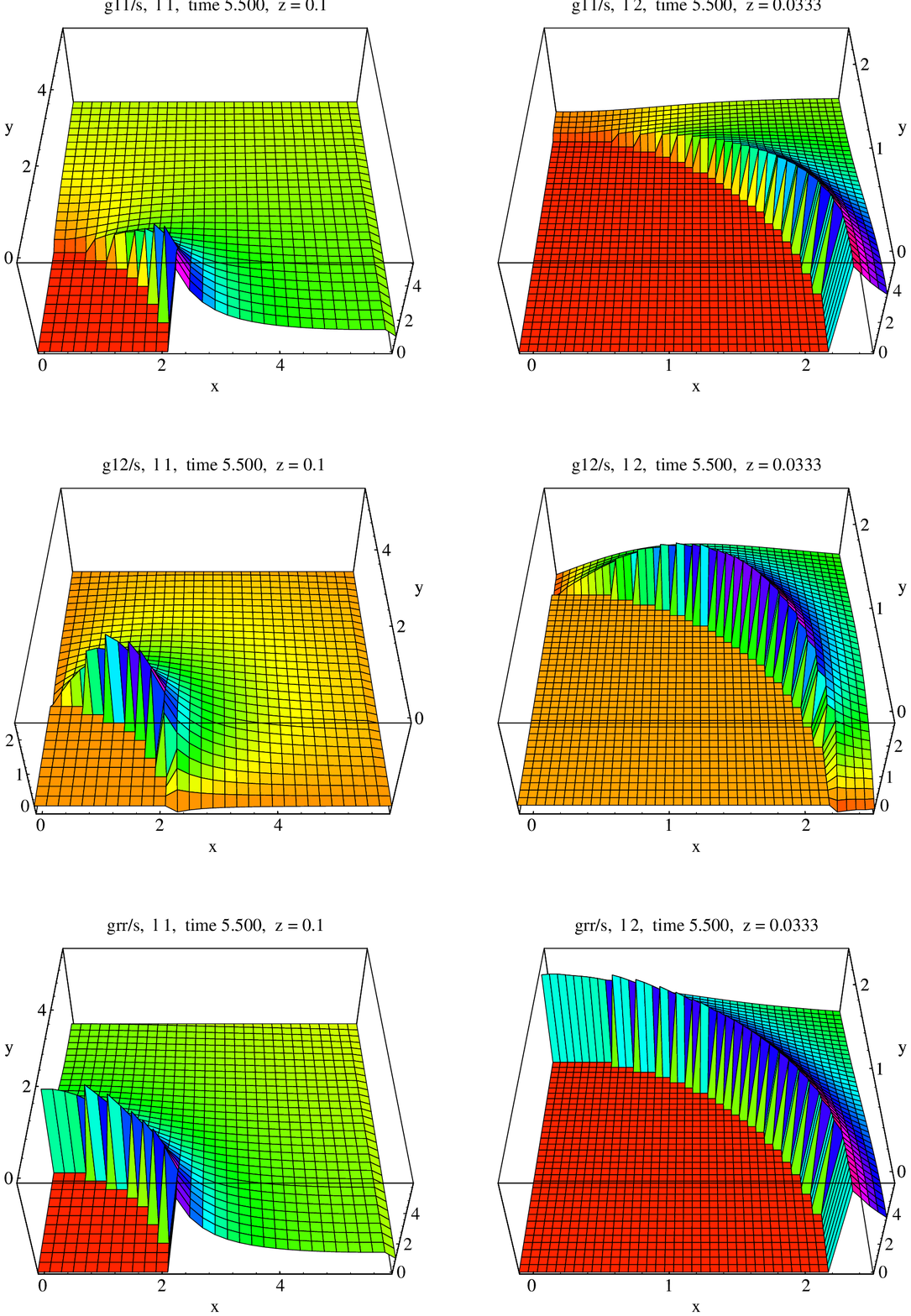}}
\caption{Metric components at levels 1 and 2 at time $\t = 5.5$.}
\label{t5}
\end{figure}

In this section we present results of our 3+1 ADM adaptive mesh code
for the Schwarzschild space time in geodesic slicing.  In figure
\ref{grr} we plot the unscaled metric component $g_{\rb\rb}$ at time
$\t$ on the horizon at radius $\rbah(\t)$. The data is taken on the
diagonal of the first octant. We set $M = 1$ in this section. Five
different data sets are plotted. The solid smooth line is the
analytical result. At time $0$, $\rbah = \fot$ and $g_{\rb\rb} = 16$
(compare (\ref{psi}) and (\ref{ginit})). Initially, there is a drop in
$g_{\rb\rb}$ as the horizon moves away from the $1/\rb$ singularity in
the conformal factor, while at late times, $\rbah \propto
\t^\frac{2}{3}$ and also $g_{\rb\rb} \propto
\t^\frac{2}{3}$, so that
$g_{\rb\rb} \propto \rbah$. These estimates are valid for $\rbah$
large enough so that $\psi(\rbah)\approx 1$, but note that at the
right edge of figure \ref{grr}, $\rbah(7) =
2.8$ and $\psi(\rbah(7))^4 = 1.9$.
The horizon moves out, but so does the $r = 0$ singularity, and it
just so happens that the horizon marks a value on the flank of the
$r=0$ singularity in the radial metric that moves to infinity
increasing linearly with the radial coordinate of the metric.

The main result of this paper is the line slightly above the analytic
curve.  It is obtained on the finest level of a three level adaptive
mesh with grid spacings 0.07, 0.21, 0.63 (refinement factor 3). The
computations are performed in the first octant with double leapfrog
and conformal differencing, and without artificial dissipation. 
At the inner boundary an apparent horizon
boundary condition is used with three buffer points, and the coarse
grids reach far enough to hold the data constant at the outer
boundary.

In figures \ref{t1} and \ref{t5} we plot for the same run two
dimensional projections of $g_{\rb\rb}$, $g_{xx}$, and $g_{xy}$ scaled
by $\psi^4$ at $\t = 1.0$ and $\t = 5.5$, respectively. The two finest
levels are shown. Level 0 does not extent further than level 1 and is
just maintained for the truncation error estimate. The data inside the
horizon minus buffer is arbitrarily set to zero since no evolution is
computed there.  Note that at $\t = 5.5$ the horizon has almost
reached the border of the level 2 grid. A small inaccuracy is visible
at the boundary of the level 1 grid at $\t = 5.5$ due to the constant
outer boundary condition. There is a corresponding deviation from zero
in the Hamiltonian constraint, which propagates through the whole
domain of integration but remains small.  Experience shows that it is
far from simple to obtain a stable evolution for the gradients in
level 1 at $\t = 5.5$. But this is one of the problems that is reduced
by adaptive mesh, since as usual the data from the level 2 grid has
been injected.

The curve that deviates wildly from the analytic solution at late
times in figure \ref{grr} belongs to a single level run under
identical conditions as defined above except that no coarser grids are
introduced. There are two further runs plotted in figure \ref{grr},
which fall just slightly below the analytic curve. They correspond to
uniform grids with spacing 0.05 and analytic data at the outer
boundary. One of the runs was performed with the Brailovskaya scheme,
which in our implementation is twice as slow as double leapfrog and
does not improve accuracy as opposed to \cite{b}.

The key limiting factor of all these runs is computer memory. Typical
runs involve one or two boxes with about $40^3$ points on a 24 Mflop
machine (linpack.c) with 80MB RAM taking 10 hours (compared to
gigaflops, gigabytes, and about the same time at NCSA
\cite{a}).  Having 40 points in any one direction is ridiculously
little compared to what is available for lower dimensional
problems. In conjunction with adaptive mesh it is clearly much more
efficient to have 2 boxes of size $40^3$ rather than one box of
$50^3$, for which the total number of points is about the same.  In
\cite{a}, for geodesic slicing a grid of size $128^3$ is used with grid
spacing 0.05 to cover about the same range of $x = 0$ to 6 as in
figures \ref{t1} and
\ref{t5}. Up to $200^3$ was managable in \cite{a}.  We find it 
surprising how well one can do with so few points per direction and a
gridspacing which by no means is `much smaller' than 1.

The apparent horizon boundary condition is working well. Even with
only about 3 points as a buffer zone, data directly at the horizon is
not significantly affected on this scale, which is apparent in figure
\ref{grr}, and which we also checked by comparing with runs for
analytic inner boundary.

There are several reasons why the runs in figure \ref{grr} cannot be
continued to later times, all of them related to size limitations. 
The truncation errors that drive the
adaptivity are spherically distributed, and given the current
resolution we do not attempt to cover spheres by several grids, so all
grids are concentric about $\rb = 0$. Hence given some maximal volume
like $40^3$, the grid spacing determins
the position of the outer boundary (see the examples below).
Referring to figure \ref{nov} for Schwarzschild in Novikov
coordinates, it is clear that at late times there is no room for the
three point buffer necessary for the apparent horizon condition. Even
before that, the steep increase in the metric coefficients makes the
evolution unstable. So, in these coordinates at this resolution we are
squeezed out at around $\t = 6$.

In \cite{a}, evolution times of around $t = 15$ to $50$ have been
obtained, which is the best one has achieved in 3+1 dimensions, but
for different coordinates (various implementations of maximal and
algebraic slicing, horizon locking shift). For the accurate extraction
of gravitational waves, on the order $t = 1000$ would be
nice. Geodesic slicing is not well suited for a code that is supposed
to run forever, because the horizon keeps moving outward and the
radial metric coefficients increase.  For the same reason, geodesic
slicing makes for an interesting test case apart from the crash test,
because one can work on some aspects of moving horizons.

The one scheme without built-in time limitations is based on horizon
locking shift conditions \cite{ADMSS}. One starts with dynamically
evolving data, but manages to find coordinates in which the metric
becomes static, which unsurprisingly is possible for the Schwarzschild
spacetime. To find the final static black hole is just what one needs
for many problems, on the other hand, our maximal proper time of $\t =
6$ is not too bad for a genuinely dynamical slicing.

\section{Conclusions and outlook}

The numerical results collected with our new 3+1 dimensional adaptive
mesh code in the case of Schwarzschild in geodesic slicing are in good
agreement with the analytical solution.  Tests of the adaptive mesh
rely mostly on no physics and flat space scenarios, but for the black
hole case the added efficiency of the adaptive mesh was crucial for
performing the computations on a small work station. The evolution
reproduces the crash time of $\pi M$, and can also be carried out to
about $6M$ with the help of an apparent horizon boundary condition.
We argued that numerical relativity on patches is a natural idea for
adaptive mesh in general relativiy.

Apart from obvious extensions of this work to bigger machines, let us
mention three directions for future work. The adaptive mesh can be
generalized to cover some aspects of numerical relativity on patches,
e.g.\ to overlapping boxes without parents. 
Having the ADM compiler available, one can experiment with the various
hyperbolic formulations that have become available recently (see
\cite{Fr} for a review).
Finally, as a simple example for non-vacuum general relativity, one
can study the collapse of a scalar field in 3+1 to find out whether
the Choptuik effect exists for non-spherical configurations of the
scalar field.

\bigskip

Acknowledgements. It is a pleasure to thank G. Allen, B. Schmidt, and
B. Schutz for helpful discussions.

\newcommand{\apny}[1]{{\em Ann.\ Phys.\ (N.Y.) }{\bf #1}}
\newcommand{\cjm}[1]{{\em Canadian\ J.\ Math.\ }{\bf #1}}
\newcommand{\cmp}[1]{{\em Commun.\ Math.\ Phys.\ }{\bf #1}}
\newcommand{\cqg}[1]{{\em Class.\ Quan.\ Grav.\ }{\bf #1}}
\newcommand{\grg}[1]{{\em Gen.\ Rel.\ Grav.\ }{\bf #1}}
\newcommand{\jgp}[1]{{\em J. Geom.\ Phys.\ }{\bf #1}}
\newcommand{\ijmp}[1]{{\em Int.\ J. Mod.\ Phys.\ }{\bf #1}}
\newcommand{\jcp}[1]{{\em J. Comp.\ Phys.\ }{\bf #1}}
\newcommand{\jmp}[1]{{\em J. Math.\ Phys.\ }{\bf #1}}
\newcommand{\mpl}[1]{{\em Mod.\ Phys.\ Lett.\ }{\bf #1}}
\newcommand{\np}[1]{{\em Nucl.\ Phys.\ }{\bf #1}}
\newcommand{\pl}[1]{{\em Phys.\ Lett.\ }{\bf #1}}
\newcommand{\pr}[1]{{\em Phys.\ Rev.\ }{\bf #1}}
\newcommand{\prl}[1]{{\em Phys.\ Rev.\ Lett.\ }{\bf #1}}
\newcommand{\bb}{B. Br\"ugmann }
\newcommand{\bib}[1]{\bibitem{#1}}

\end{document}